\documentclass[12pt]{article}
\usepackage[active]{srcltx}
\usepackage{graphicx}
\usepackage{amsmath}
\usepackage{bbold}
\usepackage{eufrak}
\usepackage{caption}
\title{\large Notes on The Feynman Checkerboard Problem}
\author{\large Keith A. Earle,$^{1,2}$ \\ 
$^{1}$ Physics Department, University at Albany\\
1400 Washington Ave, Albany NY 12222 \\
$^{2}$ Corresponding author}

\begin{document}

\maketitle

\begin{abstract}
The Feynman checkerboard problem is an interesting path integral approach to the Dirac equation in `1+1' dimensions.
I compare two approaches reported in the literature and show how they may be reconciled.
Some physical insights may be gleaned from this approach.
\end{abstract}
\vfill
\eject
\section{Introduction}\label{sec:intro}

Kauffman and Noyes\cite{KN:96} presented an intriguing derivation of the Feynman checkerboard
problem\cite{FH:65}. Of particular value, Kauffman and Noyes provide explicit expressions for
the contributions from the various equivalence classes of paths that contribute to the wave
function. Explicit evaluation of some simple cases reveals that these expressions are incorrect,
unfortunately. Prior to Kauffman and Noyes, Jacobson and Schulman\cite{JS:84} sketched a
derivation of the combinatorial factors that occur in the `RL' class of paths discussed in
Kauffman and Noyes\cite{KN:96}. By reverse engineering the arguments given in Jacobson
and Schulman\cite{JS:84}, I was able to write the correct combinatorial factors for the checkerboard
problem as developed in Kauffman and Noyes\cite{KN:96}. Section~\ref{sec:JS} is a pr\'{e}cis of
Jacobson and Schulman's constructive procedure\cite{JS:84}. Section~\ref{sec:KN} shows how to
express these results in the notation of Kauffman and Noyes\cite{KN:96}. In Section~\ref{sec:CP}, I
show how the propagator for the checkerboard problem may be evaluated for the discrete time step case
and the continuum limit. The Appendix contains a graphical representation of all of the possible paths
for the specific case of a $3 \times 2$ grid so that the combinatorial arguments in the main text may
be followed constructively.

\section{Jacobson and Schulman's Argument}\label{sec:JS}

For a rectangular grid consisting of $r$ steps of unit length in the `R' direction and $l$ steps of
unit length in the `L' direction, a `typical' `RL' path consisting of $c$ corners may be found
by noting that for an `RL' path there are exactly $1+(c-1)/2$ left turns and exactly $(c-1)/2$ right
turns where the first step must be to the right and the last step must be to the left. Adding the
number of left and right turns together, one sees that $1+(c-1)/2 +(c-1/2) = c$ as it should. Note that,
for $n$ steps in a particular direction, there are $n+1$ grid points, but the first and last grid points
are constrained by the equivalence class of the path. Thus, for $n$ steps there are $n-1$ free grid
points at which to place $(c-1)/2$ turns.  As noted above, for the `RL' paths, the last turn must be
from right to left so there are $(c-1)/2$ free turning points among $r-1$ grid points on the `R' axis
and $(c-1)/2$ free turning points among $l-1$ grid points on the `L' axis. These considerations allowed
Jacobson and Schulman\cite{JS:84} to derive the following count for the number of `RL' paths consisting
of $c$ corners
\begin{equation}
 N_{RL}(c) = \left(\begin{array}{c}
                    r-1\\ (c-1)/2
                   \end{array}\right)\left(\begin{array}{c}
                                            l-1\\ (c-1)/2
                                           \end{array}\right).\label{eq:NRLdef}
\end{equation}
This is related to the expression for $N_{RR}$ or $N_{LL}$ given by Kauffman and Noyes\cite{KN:96} with
some important differences. Note the symmetry between $l$ and $r$ in Equation~\ref{eq:NRLdef}. One may
verify by explicitly constructing the paths for small dimension grids that the expression for
$N_{RL}(c)$ in Equation~\ref{eq:NRLdef} is in fact the correct one. An example of this procedure for
a $3\times 2$ grid is given in the Appendix. Due to the symmetry in the lower
argument of the binomial coefficients, one expects that $N_{RL}(c)=N_{LR}(c)$. Explicit construction
of the allowed paths for small grids confirms this. In order to understand this result from a geometric
perspective, note that each `LR' path may be inferred from a
corresponding `RL' path by rotating a given `RL' path about the center of the grid by 180
degrees\footnote{Thanks to Kevin Knuth of the UAlbany Physics Department for pointing this out.}. The
reader can construct a few explicit examples of this procedure by considering the diagrams given in the
Appendix.

The quantity $(c-1)/2$ is the $k$ index of Kauffman
and Noyes\cite{KN:96}. In the notation of Kauffman and Noyes\cite{KN:96}
\begin{equation}
 N_{RL}(k)=N_{LR}(k) = C^{(r-1)}_kC^{(l-1)}_k\label{eq:NRLdefKN}.
\end{equation}
This expression is seen to be closer to the expressions for $N_{RR}(k)$ and $N_{LL}(k)$ given in
Kauffman and Noyes\cite{KN:96}. Equation~\ref{eq:NRLdefKN} properly counts the number of allowed paths
for both the `RL' and `LR' classes. It is related to the $\psi_0$ function of Kauffman and
Noyes\cite{KN:96}. Similar arguments may be employed to derive the appropriate expressions for
$N_{RR}(k) = C^{(r-1)}_{k+1}C^{(l-1)}_{k}$ and $N_{LL}(k) = C^{(r-1)}_{k}C^{(l-1)}_{k+1}$. There is an
error in the expressions for $N_{RR}(k)$ and $N_{LL}(k)$ in Jacobson and Schulman\cite{JS:84}, although
the correction is vanishingly small in the large grid, large number of corners limit.  The
expressions given here may be verified by explicitly counting paths for small grids.  An example of this
procedure is given in the Appendix for a $3\times 2$ grid. It is seen
that $N_{RR}(k)$ is related to the $\psi_R$ wave function and $N_{LL}(k)$ is related to the $\psi_L$ 
wave function of Kauffman and Noyes\cite{KN:96}.

\section{Kauffman and Noyes' Wave Functions}\label{sec:KN}

In order to write a correct wave function which properly accounts for the number of paths in a given
equivalence class for a grid size $r\times l$, it is only necessary to make the following identifications
based on the wavefunctions given in Kauffman and Noyes\cite{KN:96}
\begin{eqnarray}
 \psi_0 & = & \sum_{\mathrm{odd} c\geq 1} (-1)^{(c-1)/2}C^{(r-1)}_{(c-1)/2}C^{(l-1)}_{(c-1)/2}\nonumber \\
 \psi_L & = & \sum_{\mathrm{even} c > 0} (-1)^{(c/2)-1}C^{(r-1)}_{(c/2)-1}C^{(l-1)}_{(c/2)}\nonumber \\
 \psi_R & = & \sum_{\mathrm{even} c > 0} (-1)^{(c/2)-1}C^{(r-1)}_{(c/2)}C^{(l-1)}_{(c/2)-1}.\nonumber
\end{eqnarray}
Here $C^{(r-1)}_{(c-1)/2}$ for example is a `generalized binomial coefficient' in the terminology of
Kauffman and Noyes\cite{KN:96}. This generalized binomial coefficient may be written as\cite{KN:96}
$ C^{(r-1)}_{(c-1)/2} \equiv \frac{(r-1-\Delta)!}{((c-1)/2)!(r-1-\Delta-(c-1)/2)!} $.

Note that $c$ ranges over odd values for $\psi_0$ and even values for $\psi_R$ and $\psi_L$. For future
applications, it is useful to redefine the summation index so that it runs over all non-negative integers
regardless of class. For $\psi_0$, substitute $(c-1)/2\rightarrow k$. For $\psi_L$ and $\psi_R$
substitute $(c/2)-1\rightarrow k$. With this choice of summation index one finds
\begin{eqnarray}
 \psi_0 & = & \sum_{k\geq 0} (-1)^{k}\frac{(r-1)!}{k!(r-1-k)!}
                                             \frac{(l-1)!}{k!(l-1-k)!}\label{eq:p0deriv} \\
 \psi_L & = & \sum_{k\geq 0} (-1)^{k}\frac{(r-1)!}{k!(r-1-k)!}
                                             \frac{(l-1)!}{(k+1)!(l-1-(k+1))!}\label{eq:plderiv} \\
 \psi_R & = & \sum_{k\geq 0} (-1)^{k}\frac{(r-1)!}{(k+1)!(r-1-(k+1))!}
                                             \frac{(l-1)!}{k!(l-1-k)!}\label{eq:prderiv}
\end{eqnarray}
As noted above, the summation index $k$ in Equations~\ref{eq:p0deriv}--\ref{eq:prderiv} is now over all
non-negative integers regardless of path class.
With the definitions given in Equations~\ref{eq:p0deriv}--\ref{eq:prderiv}, the following derivative
identities, related to similar expressions in Kauffman and Noyes\cite{KN:96} may be written down.
These derivative identities are useful for constructing the discretized version of the `1+1' Dirac
equation.
\begin{eqnarray}
 \frac{\partial\psi_R}{\partial r} & = & \psi_0 \nonumber \\
 \frac{\partial\psi_0}{\partial r} & = & -\psi_L \nonumber \\
 \frac{\partial\psi_L}{\partial l} & = & \psi_0 \nonumber \\
 \frac{\partial\psi_0}{\partial l} & = & -\psi_R \nonumber
\end{eqnarray}
In order to compare to the one-dimensional Dirac equation, appropriate linear combinations of
Equations~\ref{eq:p0deriv}--\ref{eq:prderiv} are needed.  Note that the phase of the Dirac equation
given in Kauffman and Noyes\cite{KN:96} differs from that used in Jacobson and Schulman\cite{JS:84}.
The latter phase convention is more commonly used and will also be used here, as this leads to a
propagator of the following form\cite{JS:84}
\begin{equation}
 K_{\beta\alpha} = \lim_{n\rightarrow\infty}\sum_{c\geq 0} N_{\beta\alpha}(c)(i\epsilon m_0)^c
 \label{eq:Kdef}
\end{equation}
where $\alpha,\beta\in\{L,R\}$ and $c$ is the number of path changes and $n$ is the number of steps.
This is the form of the propagator given by Feynman in his original formulation of the checkerboard
problem\cite{FH:65}. This form is chosen for the work presented here in order to facilitate comparison
with previously published expressions. As Kauffman and Noyes note\cite{KN:96}, other choices are
possible for, \emph{e.g.}, real-valued solutions of the Dirac equation.

Using the phase convention implied by Equation~\ref{eq:Kdef}, one finds
\begin{equation}
 \left(\begin{array}{c}i\psi_2 \\ i\psi_1\end{array}\right) = 
 \left(\begin{array}{c}\partial\psi_1/\partial r \\ \partial\psi_2/\partial l\end{array}\right).
  \label{eq:Ddef}
\end{equation}
Equation~\ref{eq:Ddef} is equivalent to Equation 13 of Kauffman and Noyes \cite{KN:96} with the current
phase convention. The phase convention implied by Equation~\ref{eq:Ddef} leads to the following choices
for $\psi_1$ and $\psi_2$:
\begin{eqnarray}
 \psi_1 & = & i\psi_0-\psi_R\label{eq:psi1} \\
 \psi_2 & = & i\psi_0-\psi_L\label{eq:psi2}.
\end{eqnarray}
Equations~\ref{eq:psi1} and~\ref{eq:psi2} are correctly phased with respect to the propagator defined
by Equation~\ref{eq:Kdef}.

\section{Constructing the Propagator}\label{sec:CP}

In order to test whether the phases and wavefunctions chosen here are consistent with previously
published results, it is useful to perform a consistency check. For this reason, the propagator
for the checkerboard problem will be derived with the wavefunctions defined by
Equations~\ref{eq:p0deriv}--\ref{eq:prderiv}. In the limit as the number of corners in the path goes
to infinity the continuum limit for the propagator may be derived. Observe
(cf.\@ Equation~\ref{eq:Kdef})
\begin{equation}
 K_{RL} = K_{LR} = \lim_{n\rightarrow\infty}\sum_{c\geq 0}N_{RL}(c)(i\epsilon m_0)^{c}
\end{equation}
where $c$ is the number of corners and $m_0$ is the particle mass. Note that the path specific summation
index $c$ is used here instead of the generalized $k$ index. This choice facilitates comparison with 
Jacobson and Schulman's derivation\cite{JS:84}. In a system of units where $\hbar$ and the speed of
light are unity, $m_0$ has units of $1/\mathrm{length}$.  Recall that for `RL' or `LR' paths $c$ is odd,
$N_{RL}$ gives the number of paths connecting the endpoints of the path with $c$ corners and $n$ is the
total number of steps in the path.  It is useful to define the subsidiary quantity $m$ such that 
$r=(n+m)/2$ and $l=(n-m)/2$. Following Jacobson and Schulman's exposition\cite{JS:84}, note that
\begin{equation}
 rl = \frac{n^2}{4}\left(1-\left(\frac{m}{n}\right)^2\right).\label{eq:rlproddef}
\end{equation}
Defing the quantity $\gamma = \sqrt{1-(m/n)^2}$, the product $rl$ in Equation~\ref{eq:rlproddef} may
be written as $rl = (n/(2\gamma))^2$. The quantity $m/n$ may be interpreted as a dimensionless velocity,
as $n > 0$ and $m$ can have either sign. Note that $m=0$  corresponds to a particle at rest. If the path
difference $b-a$ is traversed in a time $t_b-t_a$ then $(m/n)^2 = (b-a)^2/(t_b-t_a)^2$ where time is
measured in light meters, say. Performing time slicing in the usual way, note that
$\epsilon = (t_b-t_a)/n$. For a large number of time steps such that
$c/n\stackrel{n\rightarrow{\infty}}{\rightarrow} 0$ the combinatorial factors in $N_{RL}(c)$ may be
evaluated asymptotically.  The steps involved are similar to the ones used to obtain a Poisson
distribution from a binomial distribution\cite{R:65}.  Note that
\begin{equation}
 N_{RL}(c) = \left(\begin{array}{c} r-1 \\ (c-1)/2 \end{array}\right)
             \left(\begin{array}{c} l-1 \\ (c-1)/2 \end{array}\right).\label{eq:rlc}
\end{equation}
When $r,l\gg c$ in Equation~\ref{eq:rlc} the ratios of factorials appearing in the binomial coefficients
may be approximated as
\begin{eqnarray}
 N_{RL}(c) & = & \frac{(r-1)!}{[(c-1)/2]![r-1-(c-1)/2]!}\frac{(l-1)!}{[(c-1)/2]![l-1-(c-1)/2]!}
 \nonumber \\
           & \approx & \frac{r^{(c-1)/2}}{[(c-1)/2]!}\frac{l^{(c-1)/2}}{[(c-1)/2]!}\nonumber \\
           & = & \frac{(lr)^{(c-1)/2}}{\left[[(c-1)/2]!\right]^2}.\label{eq:Nasym}
\end{eqnarray}
The asymptotic propagator using Equation~\ref{eq:Nasym} may thus be written
\begin{equation}
 K_{RL} = (i\epsilon m_0)\sum_c(i\epsilon m_0)^{c-1}\left(\frac{n}{2\gamma}\right)^{(c-1)}
  \frac{1}{[[(c-1)/2]!]^2}.\label{eq:KcRL}
\end{equation}
Using the definition of $\epsilon$ and introducing the abbreviation $z = m_0(t_b-t_a)/\gamma$, the
expression for the propagator may be put in to the following form
\begin{equation}
 K_{RL} = (i\epsilon m_0)\sum_{k=0}^{\infty}(-1)^k(z/2)^{2k}/(k!)^2\label{eq:Kasym}
\end{equation}
where $k= (c-1)/2$. Note that\cite{AS:64}
\begin{equation}
 J_{\alpha}(x) = 
 \sum_{s=0}^{\infty}\frac{(-1)^{s}}{s!\Gamma(s+\alpha+1)}\left(\frac{x}{2}\right)^{2s+\alpha}
 \label{eq:Jadef}
\end{equation}
where $\Gamma(s+\alpha+1) = (s+\alpha)!$ for integer $\alpha$ such that $s+\alpha \geq 0$. Using
Equation~\ref{eq:Jadef}, the asymptotic form of the $K_{RL}$ propagator may be written.
\begin{equation}
 K_{RL} = (i\epsilon m_0)J_0(z)\label{eq:KRLdef}
\end{equation}
A similar procedure can be used to evaluate the $K_{RR}$ and $K_{LL}$ contributions to the propagator.
For $K_{RR}$ paths, one has
\begin{equation}
 K_{RR} = \lim_{n\rightarrow\infty}\sum_{c}\left(\begin{array}{c}r-1 \\ c/2\end{array}\right)
                                           \left(\begin{array}{c}l-1 \\ c/2-1\end{array}\right)
                                           (i\epsilon m_0)^c\nonumber
\end{equation}
where the sum is over all even integers greater than zero.  The expression for $K_{LL}$ may be found
from the substitution $r\rightarrow l\rightarrow r$.  Using the same asymptotic analysis as before,
one finds
\begin{equation}
 K_{RR} \approx -2r\epsilon m_0\frac{\gamma}{n}\sum_{k=0}^{\infty}\frac{(-1)^k(z/2)^{2k+1}}{k!(k+1)!},
 \nonumber
\end{equation}
where the summation index $k=(c/2)-1$.
Using the definition of $J_{\alpha}$ in Equation~\ref{eq:Jadef}, this may be rewritten
\begin{equation}
 K_{RR} \approx -2r\epsilon m_0\frac{\gamma}{n} J_1(z).\nonumber
\end{equation}
Recalling that $r=(n+m)/2$ and $\gamma=1/\sqrt{1-(m/n)^2}$, and defining $t=n\epsilon$ and
$x=m\epsilon$, where $|x|<t$, one may show
\begin{equation}
 K_{RR} \rightarrow -\epsilon m_0\frac{t+x}{\tau}J_1(z),\label{eq:KRRdef}
\end{equation}
where $\tau^2 = t^2-x^2$. Making the substitution $r\rightarrow l\rightarrow r$ one may also write
down
\begin{equation}
 K_{LL} \rightarrow -\epsilon m_0\frac{t-x}{\tau}J_1(z),\label{eq:KLLdef}
\end{equation}
Noting that the number of corners is even for `RR' or `LL' paths and odd for `RL' or `LR' paths, one
may write down the continuum limit for the propagator\cite{JS:84} by dividing through by $2\epsilon$.
Combining Equations~\ref{eq:KRLdef}, \ref{eq:KRRdef} and~\ref{eq:KLLdef} one may write down the
continuum propagator in a concise matrix form as follows
\begin{equation}
 K = \left(\begin{array}{cc}
            K_{RR} & K_{RL} \\
            K_{LR} & K_{LL}
           \end{array}
\right) = \frac{m_0}{2}\left(\begin{array}{cc} -\frac{t+x}{\tau}J_1(z) & iJ_0(z) \\
                                          iJ_0(z) & -\frac{t-x}{\tau}J_1(z) \end{array}\right)
\label{eq:KCdef}
\end{equation}
Equation~\ref{eq:KCdef} reproduces Jacobson and Schulman's result\cite{JS:84}. This form allows efficient
computation of the two-component wave function at time $t_b$ given the value of the wavefunction at time
$t_a$. Recall that $z\equiv m_0(t_b-t_a)/\gamma$.  Specifically,
\begin{equation}
 \left(\begin{array}{c}
        \psi_R(t_b)\\
	\psi_L(t_b)
       \end{array}
\right) = \frac{m_0}{2}\left(\begin{array}{cc} -\frac{t+x}{\tau}J_1(z) & iJ_0(z) \\
                                          iJ_0(z) & -\frac{t-x}{\tau}J_1(z) \end{array}\right)
\left(\begin{array}{c}
       \psi_R(t_a)\\
	\psi_L(t_b)
      \end{array}
\right)
\label{eq:Psitbdef}
\end{equation}
With this construction, it is clear why the propagator $K$ is decomposed into $K_{RR}$, $K_{RL}$,
$K_{LR}$ and $K_{LL}$ terms.

\section{Estimating the Number of Significant Paths}

From the discussion given in the appendix, it is clear that the number of paths with a given number of
corners is not uniform. On a fine grid, it is useful to develop a criterion for the paths with a given
number of corners that make the most significant contributions to the propagator $K$. The discussion
given here parallels that of Jacobson and Schulman~\cite{JS:84} with the combinatorial factors worked
out here.  Equation~\ref{eq:KcRL} is a useful starting point for discussion.  Using the definition of
$z$, Equation~\ref{eq:KcRL} can be rewritten as follows
\begin{equation}
 K_{RL}\approx\frac{2\gamma}{n}\sum_{\mbox{odd $c$}}i^{c}(z/2)^c/[[(c-1)/2]!]^2.\label{eq:KzcRL}
\end{equation}
We seek that value of $c$ which maximizes $(z/2)^c/[[(c-1)/2]!]^2$. Following Jacobson and Schulman
\cite{JS:84}, we define $\exp f(c) = (z/2)^c/[[(c-1)/2]!]^2$. Thus
\begin{equation}
 f(c) = c\log(z/2) - 2\log([(c-1)/2]!).\label{eq:fcdef}
\end{equation}
Using Stirling's approximation, Equation~\ref{eq:fcdef} becomes
\begin{equation}
 f(c) \approx c\log(z/2) -2[(c-1)/2]\log((c-1)/2) + 2[(c-1)/2] - \cdots
\end{equation}
Setting $df/dc = 0$ we find $c_0 = z$ where $c_0$ is that value of $c$ which extremizes $\exp f(c)$. In
order to verify that $c_0$ maximizes $f(c)$, compute $d^2f/dc^2 = -1/c_0$.  In the neighborhood of $c_0$,
therefore
\begin{equation}
 \exp f(c) \approx \exp f(c_0)\exp\left[-\frac{(c-c_0)^2}{2c_0}\right].\label{eq:fcasym}
\end{equation}
Thus the important contributions to $K_{RL}$ come from a narrow range of $c$ values centered on
$c_0 = z$. Similar calculations may be done for $K_{RR}$ and $K_{LL}$. The quantity $K_{LR} = K_{RL}$.
The results are
\begin{eqnarray}
 K_{RR} & = & r\sum_{\mbox{even $c > 0$}}(i)^c2^z\frac{z!}{(z/2)!\,(z/2)!}
  \exp\left[-\frac{(c-z)^2}{2z}\right] \nonumber \\
 K_{RL} & = & \frac{2\gamma}{n}\sum_{\mbox{odd $c > 0$}}(i)^c2^z\frac{z!}{(z/2)!\,(z/2)!}
  \exp\left[-\frac{(c-z)^2}{2z}\right] \nonumber \\
 K_{LR} & = & K_{RL} \nonumber \\
 K_{LL} & = & l\sum_{\mbox{even $c > 0$}}(i)^c2^z\frac{z!}{(z/2)!\,(z/2)!}
  \exp\left[-\frac{(c-z)^2}{2z}\right] \nonumber
\end{eqnarray}
Note that the summands in this limit differ only by a phase. The work of Ord and collaborators 
\cite{O:92a,O:92b,OM:93} has investigated the implications of the cyclic nature of the propagator phase
and have used it to suggest a way of extending the Feynman checkerboard problem to higher dimensions.
It is also interesting to note that there is an alternative method of evaluating the propagator, based
on the Ising model as discussed by Gersch \cite{G:81}. Schulman \cite{S:05} has references to the papers
of Ord and coworkers in addition to alternative approaches to extending the checkerboard model to higher
dimensions.

\section*{Acknowledgements}

KAE thanks the University at Albany for partial support of this work through its Faculty Research Award
Program. KAE also thanks ACERT (NIH NCRR P41 RR016292) for the use of its computational resources.

\bibliography{Earle_docs,Intrinsic,%
Analytical,Dirac,Statistical}

\begin{thebibliography}{10}

\bibitem{AS:64}
M.~Abramowitz and I.~Stegun.
\newblock {\em Handbook of Mathematical Functions with Formulas, Graphs, and
  Mathematical Tables}.
\newblock Dover, 1964.

\bibitem{FH:65}
Richard~P. Feynman and A.~R. Hibbs.
\newblock {\em {Quantum Mechanics and Integrals}}.
\newblock McGraw-Hill, 1965.

\bibitem{G:81}
H.~A. Gersch.
\newblock {Feynman's Relativistic Chessboard as an Ising Model}.
\newblock {\em International Journal of Theoretical Physics}, 20:491--501,
  1981.

\bibitem{JS:84}
Theodore Jacobson and Lawrence~S. Schulman.
\newblock {Quantum stochastics: the passage from a relativistic to a
  non-relativistic path integral}.
\newblock {\em J. Phys. A: Math. Gen.}, 17:375--383, 1984.

\bibitem{KN:96}
Louis~H. Kauffman and H.~Pierre Noyes.
\newblock {Discrete physics and the Dirac equation}.
\newblock {\em Physics Letters}, A218:139--146, 1996.

\bibitem{O:92a}
G.~N. Ord.
\newblock {A Reformulation of the Feynman Chessboard Model}.
\newblock {\em Journal of Statistical Physics}, 66:647--659, 1992.

\bibitem{O:92b}
G.~N. Ord.
\newblock {Classical Analogue of Quantum Phase}.
\newblock {\em International Journal of Theoretical Physics}, 31:1177--1195,
  1992.

\bibitem{OM:93}
G.~N. Ord and D.~G.~C. McKeon.
\newblock {On the Dirac Equation in 3+1 Dimensions}.
\newblock {\em Annals of Physics}, 222:244--253, 1993.

\bibitem{R:65}
F.~Reif.
\newblock {\em Foundations of Statistical and Thermal Physics}.
\newblock McGraw-Hill, 1965.

\bibitem{S:05}
L.~S. Schulman.
\newblock {\em {Techniques and Applications of Path Integration}}.
\newblock Dover, 2005.

\end{thebibliography}
\bibliographystyle{plain}

\appendix
\vfill
\eject

\section{Explicit Construction on a 3$\times$2 Grid}

In order to get a feel for the practical aspects of the combinatorial factors used here, consider the
problem of enumerating all allowed paths with one, two, three or four corners on a $3\times 2$ grid.
Note that there are $^{(R+L)}C_{R} = \left(\begin{array}{c}
                                            R+L\\
					      R
                                           \end{array}
\right)$ total paths\footnote{I am indebted to Prof.\ Kevin Knuth of the University at Albany Physics
Department for this observation}. With $R=3$ and $L=2$ there are $5!/(3!\,2!) = 10$ possible paths. Here
are, respectively, the `RL' and 'LR' paths with one corner\bigskip

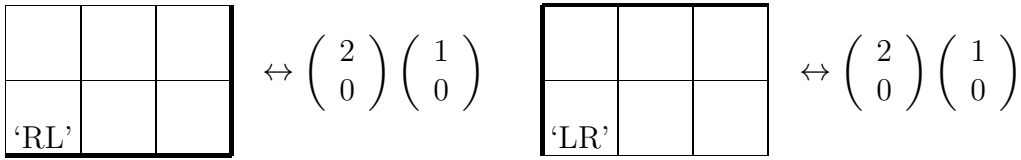
\setlength{\unitlength}{10mm}
\begin{picture}(7, 2)
  \linethickness{0.075mm}
  \multiput(0, 0)(1, 0){4}{\line(0, 1){2}}
  \multiput(0, 0)(0, 1){3}{\line(1, 0){3}}
% \linethickness{0.15mm}
% \multiput(0, 0)(5, 0){7}{\line(0, 1){20}}
% \multiput(0, 0)(0, 5){5}{\line(1, 0){30}}
  \linethickness{0.45mm}
  \put(0,0){\line(1,0){3}}
  \put(3,0){\line(0,1){2}}
  \put(3,1){\quad$\leftrightarrow \left(\begin{array}{c}
        2 \\ 0
       \end{array}
\right)\left(\begin{array}{c}
        1 \\ 0
       \end{array}
\right) $}
  \put(0.1,0.1){`RL'}
% \multiput(0, 5)(0, 10){2}{\line(1, 0){30}}  
\end{picture} 
\begin{picture}(6, 2)
  \linethickness{0.075mm}
  \multiput(0, 0)(1, 0){4}{\line(0, 1){2}}
  \multiput(0, 0)(0, 1){3}{\line(1, 0){3}}
% \linethickness{0.15mm}
% \multiput(0, 0)(5, 0){7}{\line(0, 1){20}}
% \multiput(0, 0)(0, 5){5}{\line(1, 0){30}}
  \linethickness{0.45mm}
  \put(0,0){\line(0,1){2}}
  \put(0,2){\line(1,0){3}}
  \put(3,1){$\quad\leftrightarrow \left(\begin{array}{c}
        2 \\ 0
       \end{array}
\right)\left(\begin{array}{c}
        1 \\ 0
       \end{array}
\right) $}
% \multiput(0, 5)(0, 10){2}{\line(1, 0){30}}  
  \put(0.1,0.1){`LR'}
\end{picture}

Here are the `RR' paths with two corners\bigskip

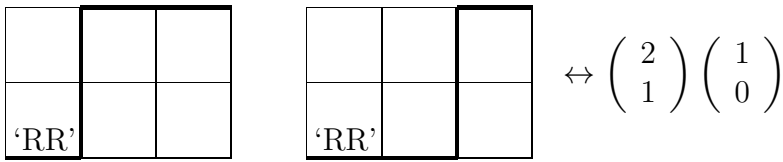
\begin{picture}(9, 2)
  \linethickness{0.075mm}
  \multiput(0, 0)(1, 0){4}{\line(0, 1){2}}
  \multiput(0, 0)(0, 1){3}{\line(1, 0){3}}
% \linethickness{0.15mm}
% \multiput(0, 0)(5, 0){7}{\line(0, 1){20}}
% \multiput(0, 0)(0, 5){5}{\line(1, 0){30}}
  \linethickness{0.45mm}
  \put(0,0){\line(1,0){1}}
  \put(1,0){\line(0,1){2}}
  \linethickness{0.45mm}
  \put(1,2){\line(1,0){2}}
  \put(0.1,0.1){`RR'}
% \multiput(0, 5)(0, 10){2}{\line(1, 0){30}}  
  \linethickness{0.075mm}
  \multiput(4, 0)(1, 0){4}{\line(0, 1){2}}
  \multiput(4, 0)(0, 1){3}{\line(1, 0){3}}
% \linethickness{0.15mm}
% \multiput(0, 0)(5, 0){7}{\line(0, 1){20}}
% \multiput(0, 0)(0, 5){5}{\line(1, 0){30}}
  \linethickness{0.45mm}
  \put(4,0){\line(1,0){2}}
  \put(6,0){\line(0,1){2}}
  \linethickness{0.45mm}
  \put(6,2){\line(1,0){1}}
% \multiput(0, 5)(0, 10){2}{\line(1, 0){30}}  
  \put(7,1){$\quad\leftrightarrow \left(\begin{array}{c}
        2 \\ 1
       \end{array}
\right)\left(\begin{array}{c}
        1 \\ 0
       \end{array}
\right) $}
  \put(4.1,0.1){`RR'}
\end{picture}

Here is the unique `LL' path with 2 corners\bigskip

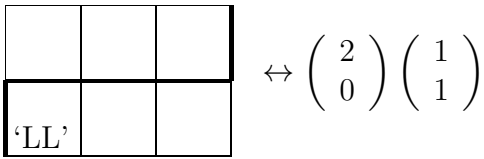
\begin{picture}(6,2)
  \linethickness{0.075mm}
  \multiput(0, 0)(1, 0){4}{\line(0, 1){2}}
  \multiput(0, 0)(0, 1){3}{\line(1, 0){3}}
% \linethickness{0.15mm}
% \multiput(0, 0)(5, 0){7}{\line(0, 1){20}}
% \multiput(0, 0)(0, 5){5}{\line(1, 0){30}}
  \linethickness{0.45mm}
  \put(0,0){\line(0,1){1}}
  \put(0,1){\line(1,0){3}}
  \linethickness{0.45mm}
  \put(3,1){\line(0,1){1}}
  \put(0.1,0.1){`LL'}
  \put(3,1){$\quad\leftrightarrow \left(\begin{array}{c}
        2 \\ 0
       \end{array}
\right)\left(\begin{array}{c}
        1 \\ 1
       \end{array}
\right) $}
\end{picture}

Here are the `RL' paths with 3 corners\bigskip

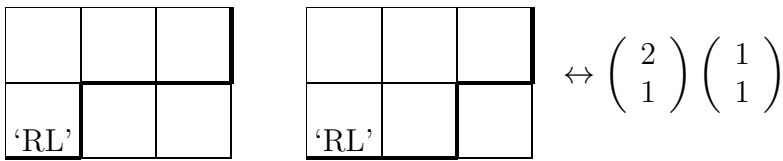
\begin{picture}(9, 2)
  \linethickness{0.075mm}
  \multiput(0, 0)(1, 0){4}{\line(0, 1){2}}
  \multiput(0, 0)(0, 1){3}{\line(1, 0){3}}
% \linethickness{0.15mm}
% \multiput(0, 0)(5, 0){7}{\line(0, 1){20}}
% \multiput(0, 0)(0, 5){5}{\line(1, 0){30}}
  \linethickness{0.45mm}
  \put(0,0){\line(1,0){1}}
  \put(1,0){\line(0,1){1}}
  \linethickness{0.45mm}
  \put(1,1){\line(1,0){2}}
  \put(3,1){\line(0,1){1}}
  \put(0.1,0.1){`RL'}
% \multiput(0, 5)(0, 10){2}{\line(1, 0){30}}  
  \linethickness{0.075mm}
  \multiput(4, 0)(1, 0){4}{\line(0, 1){2}}
  \multiput(4, 0)(0, 1){3}{\line(1, 0){3}}
% \linethickness{0.15mm}
% \multiput(0, 0)(5, 0){7}{\line(0, 1){20}}
% \multiput(0, 0)(0, 5){5}{\line(1, 0){30}}
  \linethickness{0.45mm}
  \put(4,0){\line(1,0){2}}
  \put(6,0){\line(0,1){1}}
  \put(6,1){\line(1,0){1}}
  \put(7,1){\line(0,1){1}}
% \multiput(0, 5)(0, 10){2}{\line(1, 0){30}}  
  \put(7,1){$\quad\leftrightarrow \left(\begin{array}{c}
        2 \\ 1
       \end{array}
\right)\left(\begin{array}{c}
        1 \\ 1
       \end{array}
\right) $}
  \put(4.1,0.1){`RL'}
\end{picture}

Here are `LR' paths with three corners\bigskip

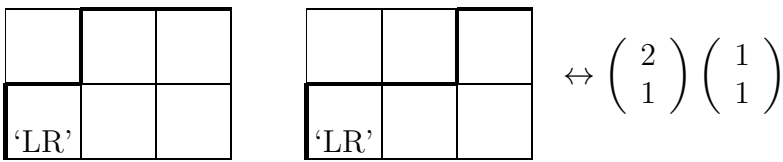
\begin{picture}(6, 2)
  \linethickness{0.075mm}
  \multiput(0, 0)(1, 0){4}{\line(0, 1){2}}
  \multiput(0, 0)(0, 1){3}{\line(1, 0){3}}
% \linethickness{0.15mm}
% \multiput(0, 0)(5, 0){7}{\line(0, 1){20}}
% \multiput(0, 0)(0, 5){5}{\line(1, 0){30}}
  \linethickness{0.45mm}
  \put(0,0){\line(0,1){1}}
  \put(0,1){\line(1,0){1}}
  \linethickness{0.45mm}
  \put(1,1){\line(0,1){1}}
  \put(1,2){\line(1,0){2}}
  \put(0.1,0.1){`LR'}
% \multiput(0, 5)(0, 10){2}{\line(1, 0){30}}  
  \linethickness{0.075mm}
  \multiput(4, 0)(1, 0){4}{\line(0, 1){2}}
  \multiput(4, 0)(0, 1){3}{\line(1, 0){3}}
% \linethickness{0.15mm}
% \multiput(0, 0)(5, 0){7}{\line(0, 1){20}}
% \multiput(0, 0)(0, 5){5}{\line(1, 0){30}}
  \linethickness{0.45mm}
  \put(4,0){\line(0,1){1}}
  \put(4,1){\line(1,0){2}}
  \put(6,1){\line(0,1){1}}
  \put(6,2){\line(1,0){1}}
% \multiput(0, 5)(0, 10){2}{\line(1, 0){30}}  
  \put(7,1){$\quad\leftrightarrow \left(\begin{array}{c}
        2 \\ 1
       \end{array}
\right)\left(\begin{array}{c}
        1 \\ 1
       \end{array}
\right) $}
  \put(4.1,0.1){`LR'}
\end{picture}

Here is the unique RR path with 4 corneres
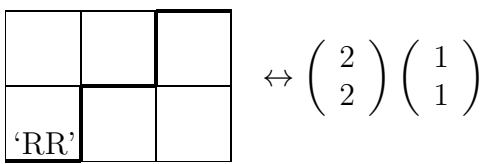
\bigskip

\begin{picture}(6, 2)
  \linethickness{0.075mm}
  \multiput(0, 0)(1, 0){4}{\line(0, 1){2}}
  \multiput(0, 0)(0, 1){3}{\line(1, 0){3}}
% \linethickness{0.15mm}
% \multiput(0, 0)(5, 0){7}{\line(0, 1){20}}
% \multiput(0, 0)(0, 5){5}{\line(1, 0){30}}
  \linethickness{0.45mm}
  \put(0,0){\line(1,0){1}}
  \put(1,0){\line(0,1){1}}
  \linethickness{0.45mm}
  \put(1,1){\line(1,0){1}}
  \put(2,1){\line(0,1){1}}
  \put(2,2){\line(1,0){1}}
  \put(0.1,0.1){`RR'}
  \put(3,1){$\quad\leftrightarrow \left(\begin{array}{c}
        2 \\ 2
       \end{array}
\right)\left(\begin{array}{c}
        1 \\ 1
       \end{array}
\right) $}
\end{picture}

Note that this exhausts all the allowed paths on a $3\times 2$ grid.
\end{document}